\documentclass[aps,prapplied,reprint,superscriptaddress,longbibliography]{revtex4-1}
\usepackage[english]{babel}
\usepackage[utf8]{inputenc}
\usepackage{amsthm}
\usepackage{mathtools}
\usepackage{physics}
\usepackage{xcolor}
\usepackage{graphicx}
\usepackage[left=23mm,right=13mm,top=35mm,columnsep=20pt]{geometry} 
\usepackage{adjustbox}
\usepackage{placeins}
\usepackage[T1]{fontenc}
\usepackage{csquotes}
\usepackage[pdftex, pdftitle={Article}, pdfauthor={Author}]{hyperref} 

\emergencystretch=\maxdimen
\begin{document}


\title{Diffusive and Fluidlike Motion of Homochiral Domain Walls in\\Easy-Plane Magnetic Strips}

\author{David A. Smith}
    \email[Correspondence email address: ]{smithd22@vt.edu}
    \affiliation{Department of Physics, Virginia Tech, Blacksburg, VA, 24061, U.S.A.}

\author{So Takei}
    \email[Correspondence email address: ]{So.Takei@qc.cuny.edu}
    \affiliation{Department of Physics, Queens College of the City University of New York, Queens, New York, 11367, U.S.A.}
    \affiliation{Physics Doctoral Program, The Graduate Center of the City University of New York, New York, New York 10016, U.S.A.}

\author{Bella Brann}
    \affiliation{Department of Physics, Virginia Tech, Blacksburg, VA, 24061, U.S.A.}

\author{Lia Compton}
    \affiliation{Department of Physics, Virginia Tech, Blacksburg, VA, 24061, U.S.A.}

\author{Fernando Ramos-Diaz}
    \affiliation{Department of Physics, Virginia Tech, Blacksburg, VA, 24061, U.S.A.}
    
\author{Matthew Simmers}
    \affiliation{Academy of Integrated Science, Virginia Tech, Blacksburg, VA, 24061, U.S.A.}

\author{Satoru Emori}
    \email[Correspondence email address: ]{semori@vt.edu}
    \affiliation{Department of Physics, Virginia Tech, Blacksburg, VA, 24061, U.S.A.}

\date{\today} 


\begin{abstract}
Propagation of easy-plane magnetic precession can enable more efficient spin transport than conventional spin waves. Such easy-plane spin transport is typically understood in terms of a hydrodynamic model, partially analogous to superfluids. Here, using micromagnetic simulations, we examine easy-plane spin transport in magnetic strips as the motion of a train of domain walls rather than as a hydrodynamic flow. We observe that the motion transitions from diffusive to fluid-like as the density of domain walls is increased. This transition is most evident in notched nanostrips, where the the domain walls are pinned by the notch defect in the diffusive regime but propagate essentially unimpeded in the fluid-like regime. Our findings suggest that spin transport via easy-plane precession, robust against defects, is achievable in strips based on realistic metallic ferromagnets and hence amenable to practical device applications.
\end{abstract}



\maketitle


\section{Introduction}

\begin{figure}[t]
    \centering
    \includegraphics[width=\columnwidth,trim= 0.75in 4.5in 0.75in 0in]{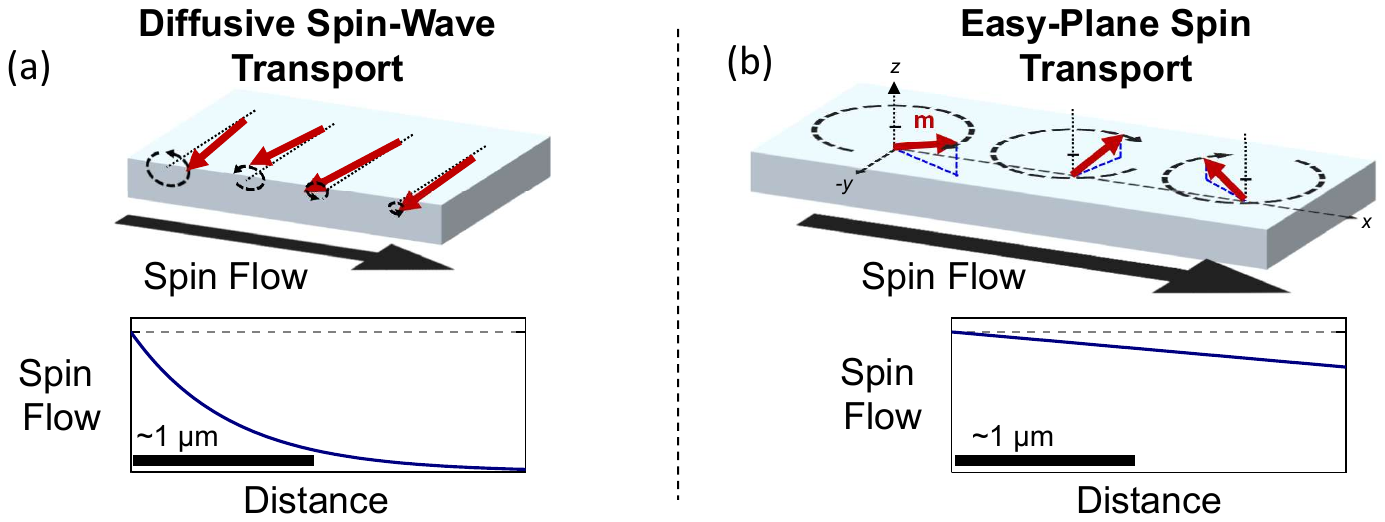}
    \caption{(a) Illustration of small angle precession constituting diffusive spin waves and exponential decay of spin flow. (b) Easy-plane precession constituting superfluid-like spin transport and associated linear decay of spin flow.}
    \label{fig:comparison}
\end{figure}

Transport of spin information via magnetization dynamics is a key area of rapid development within spintronics \cite{Sander2017}. To date, much work on micron-scale spin transport has focused on using diffusive spin waves \cite{Lebrun2018,Giles2015}. The magnetization precession cone angle in diffusive spin waves is typically $\ll 10^\circ$, and the associated spin flow decays exponentially with decay length inversely proportional to the Gilbert damping parameter $\alpha$, as illustrated in Fig.~\ref{fig:comparison}(a). As a result, efficient spin transport at or beyond the micron scale has been difficult to attain, particularly in typical metallic ferromagnets with $\alpha > 10^{-3}$ that are compatible with industrial device fabrication.

An alternative method to achieving long distance spin transport in the form of spin superfluidity \cite{Sonin2010,Chen2014,Takei2014,Skarsvag2015,Tserkovnyak2018,Iacocca2017PRBa,Schneider2021} has gathered interest in recent years. In spin superfluidity the magnetization undergoes easy-plane precession with a cone angle of $\approx 90^\circ$, driven by a current-induced spin-transfer torque~\cite{Brataas2012, Slonczewski1996, Houssameddine2007}. The resulting precessional dynamics propagates along the ferromagnet in a spiraling manner, as illustrated in Fig.~\ref{fig:comparison}(b), and is protected from unwinding by the strong easy-plane anisotropy preventing phase slips \cite{Kim2016}. While true superfluidity (i.e. lossless spin transport) is not possible as a result of ever-present viscous Gilbert damping, this unique form of magnetization dynamics creates a spin flow that decays linearly or algebraically with distance. This easy-plane superfluid-\emph{like} spin transport -- also called ``dissipative exchange flow'' \cite{Iacocca2017PRBa} or ``exchange-mediated spin transport'' \cite{Schneider2021} -- has been proposed as a means of spin information transport even in metallic ferromagnets \cite{Chen2014,Konig2001,Iacocca2017PRBa,Iacocca2017PRL,Iacocca2019} with moderate damping parameters.

Halperin and Hohenberg originally proposed a model to view easy-plane precessional magnetization dynamics from a hydrodynamic perspective \cite{Halperin1969}, in a manner that is analogous to that of superfluidity. This hydrodynamic perspective has been used to analyze easy-plane spin transport in several studies \cite{Iacocca2017PRBa,Iacocca2017PRL,Iacocca2019,Iacocca2020}. However, these studies have focused on the regime that requires higher drive current densities, $J_c$. The requirement of high current densities ($J_c > 1 \times 10^{12}$ A/m$^2$) poses potential problems in the form of Joule heating as well as electromigration altering material properties. While studies have investigated the effects of in-plane magnetocrystalline anisotropy \cite{Iacocca2017PRBa}, Gilbert damping \cite{Iacocca2019}, and void defects \cite{Iacocca2017PRL,Iacocca2020}, how the easy-plane spin transport behaves at lower drive current densities, closer to the range of experimental feasibility, has yet to be answered.

In this study, we have performed micromagnetic simulations of easy-plane spin transport in synthetic antiferromagnet nanostrips, focusing on the low drive regime. The synthetic antiferromagnet material parameters mimic those of experimentally measured, metallic ferromagnets. Instead of taking the conventional approach from a hydrodynamic perspective, we study the dynamics as a train of interacting, homochiral domain walls (DWs) \cite{Kim2017a}. We find that at low drive current densities $J_c$, the DWs can be pinned by a notch defect. We observe the transition from diffusive motion to fluid-like motion as $J_c$ is increased and the DW density increases. The dynamics of the DW train converges to that of the established hydrodynamic behavior when the DW spacing becomes comparable to the DW width at $J_c \simeq 5 \times 10^{11}$ A/m$^2$. In this fluid-like regime, the train of DWs are unimpeded by the notch defect. Our results suggest that even at moderately low $J_c$ and with deep notch defects, it is feasible to achieve easy-plane spin transport in a metallic ferromagnetic system.


\section{Simulation Details}

We have simulated easy-plane spin transport -- i.e., motion of a train of spiraling homochiral transverse N\'eel DWs -- in magnetic nanostrips using Mumax$^3$, an open-source GPU accelerated micromagnetic simulation package \cite{Vansteenkiste2014}. In single-layer ferromagnetic strips (see Appendix \ref{app:singlelayer}), the moving transverse DWs are unstable and transform into vortex DWs \cite{Beach2005,Mougin2007}, which effectively constitute phase slips and breakdown of coherent easy-plane spin transport. We instead focus here on simulations of  synthetic antiferromagnetic strips, which are composed of two ferromagnetic layers coupled in an  antiparallel manner \cite{Duine2018}. The interlayer-coupled magnetic moments reduce dipolar fields at the strip edges via flux closure and stabilize transverse N\'eel DWs \cite{Lepadatu2017}. Thus, the formation of vortices are suppressed and easy-plane spin transport, carried by spiraling transverse DWs, remains far more stable in synthetic antiferromagnets than in single-layer ferromagnets. The enhanced stability of easy-plane spin transport in synthetic antiferromagnets has been previously reported in a micromagnetic study by Skarsv\aa g \textit{et al.}~\cite{Skarsvag2015}.

\begin{figure}[t]
    \centering
    \includegraphics[width=\columnwidth,trim=0in 1.5in 0in 0.5in]{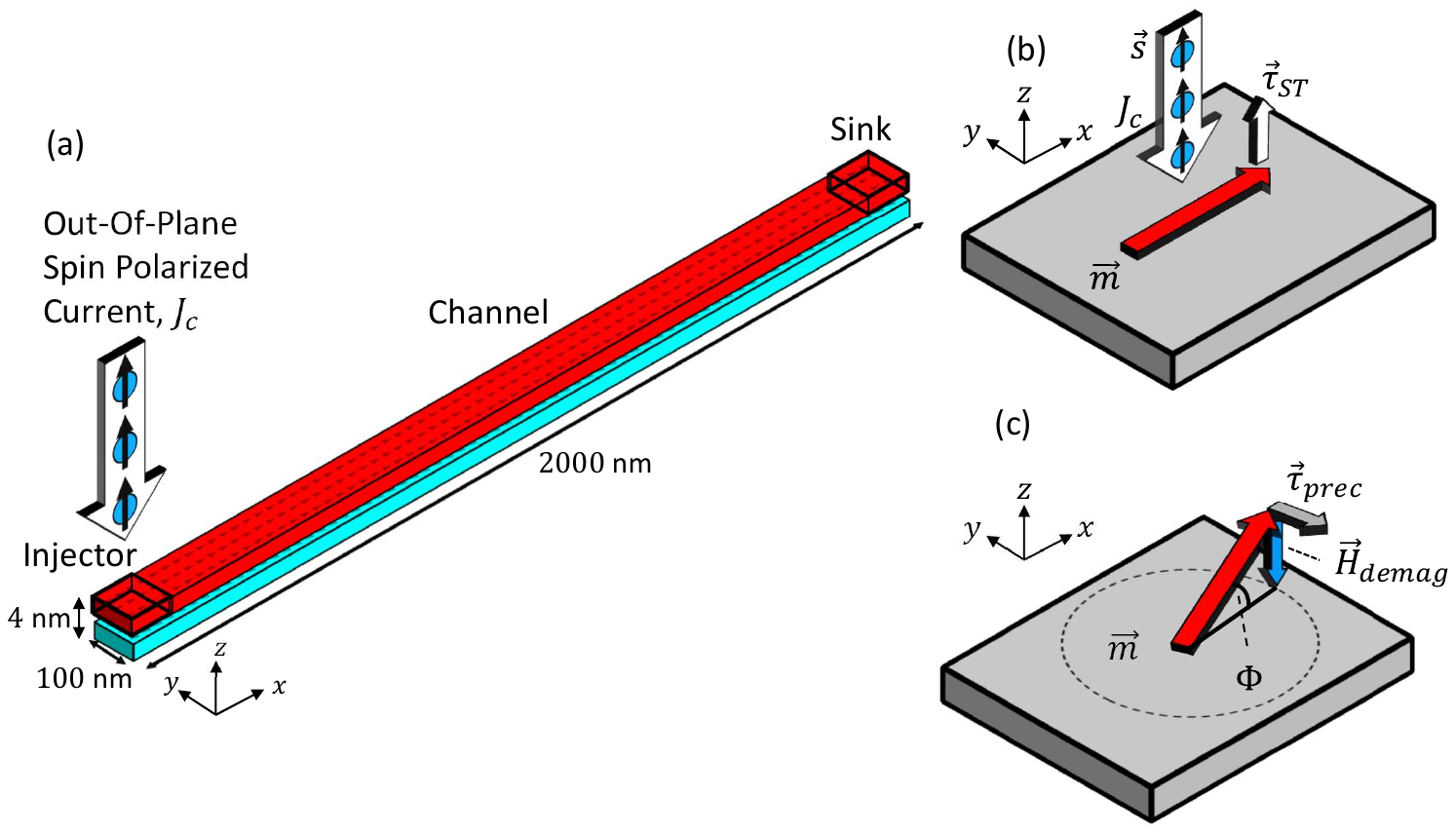}
    \caption{(a) Micromagnetic simulation setup of the synthetic antiferromagnet nanostrip. (b) The resulting torque generated by the out-of-plane spin-polarized electric current $J_c$, lifting the magnetization out of the plane in the injector region. (c) The out-of-plane component of the magnetization creates a demagnetizing field, generating a precessional torque that drives easy-plane precession.}
    \label{fig:SimulationSetUp}
\end{figure}

A depiction of our simulation set-up is shown in Fig.~\ref{fig:SimulationSetUp}(a). The dimensions of an individual ferromagnetic layer are $2000$ nm $\times$ $100$ nm $\times$ $2$ nm with a cell size of $2.5$ nm $\times$ $2.5$ nm $\times$ $2$ nm. The two layers are coupled using an RKKY interaction with strength $J_{RKKY} = -1$ mJ/m$^2$. The initial magnetization states lie completely in plane and are parallel to the long axis of the nanostrip (i.e. $\vec{m}_i \parallel \pm \hat{x}$). To simulate the interaction of easy-plane spin transport with defects, a pair of symmetric, triangular notches with lateral dimensions $60$ nm $\times$ $30$ nm were introduced at the midpoint of the nanostrip ($x = 1000$ nm).

The material parameters of our nanostrips were chosen to match those of experimentally measured, $2$ nm thick polycrytalline Fe$_{80}$V$_{20}$ (see Appendix \ref{app:PolyFeV} for determination of material parameters): saturation magnetization $M_{sat} = 720$ kA/m, in-plane magnetocrystalline anisotropy $K = 0$ J/m$^3$, and Gilbert damping parameter $\alpha = 0.006$. The exchange constant was set to $A_{ex} = 20$ pJ/m, in line with typical literature values for Fe \cite{Ma2011,Niitsu2020}. At each end of the nanostrip in a $100$ nm $\times$ $100$ nm region, we introduce an enhancement to the Gilbert damping parameter, $\alpha' = 0.015$, to simulate the effects of spin pumping into and out of the nanostrip \cite{Berger2018}. The total Gilbert damping parameter in these end regions is $\alpha_{total} = \alpha + \alpha'$. All simulations were performed at zero temperature.

In order to excite dynamics, an out-of-plane spin polarized charge current density $J_c$ was applied to the injection region, as shown in Fig. \ref{fig:SimulationSetUp}(a). The spin polarized charge current imparts an out-of-plane spin-transfer torque \cite{Brataas2012} $\vec{\tau}_{ST} \sim \vec{m} \times (\vec{s} \times \vec{m})$, where $\vec{s} \parallel \hat{z}$ is the spin polarization, on the magnetization $\vec{m}$. This excitation is similar to that in current-perpendicular-to-plane perpendicularly magnetized spin valves \cite{Slonczewski1996, Houssameddine2007}. The spin-transfer torque was set to act directly on the top ferromagnetic layer only. This was done to be consistent with previous studies \cite{Ghosh2012,Lim2021} showing that injected spins orthogonal to $\vec{m}$ in a metallic ferromagnet are absorbed within the first $\approx$1 nm. The spin polarization of the current was set to $P = 0.5$. 

The spin-transfer torque creates a finite out-of-plane component of the magnetization, $m_z$, with an out-of-plane canting angle $\Phi$, shown in Fig.~\ref{fig:SimulationSetUp}(b). The out-of-plane component $m_z$ generates a demagnetizing field $\vec{H}_{demag}$ and a precessional torque $\vec{\tau}_{prec} \sim -\vec{m} \times \vec{H}_{demag}$, as depicted in Fig.~\ref{fig:SimulationSetUp}(c). The torque then causes $\vec{m}$ to rotate in a constant direction (e.g. clockwise in the present case) and thus dictates the chirality of the resulting DWs. The easy-plane magnetization dynamics then propagates along the nanostrip, away from the injector, via exchange coupling.


\section{Results and Discussion}


\subsection{Diffusive Motion of an Isolated Domain  Wall}\label{sec:diffusivedw}

In this section, we discuss the behavior of an isolated DW in both the perfect and notched nanostrips. Both simulations were performed identically at a charge current density of $J_c = 2.4 \times 10^{11}$ A/m$^2$. In order to rotate the magnetization, the energy supplied by the current-induced spin-transfer torque must overcome the energy barrier from the uniaxial shape anisotropy of the nanostrip. This implies a threshold current density required to excite the dynamics, i.e., inject a DW into the channel. Additionally when the drive current density is sufficiently low, only a single DW can be injected into the nanostrip. When the magnetization is rotated by $180^\circ$, a $180^\circ$ DW is created at the boundary of the source. The DW is then injected into the nanostrip and driven by the out-of-plane canting angle $\Phi$. 

\begin{figure*}[ht]
    \centering
    \includegraphics[width=\textwidth,trim = 0in 0in 0in 1in]{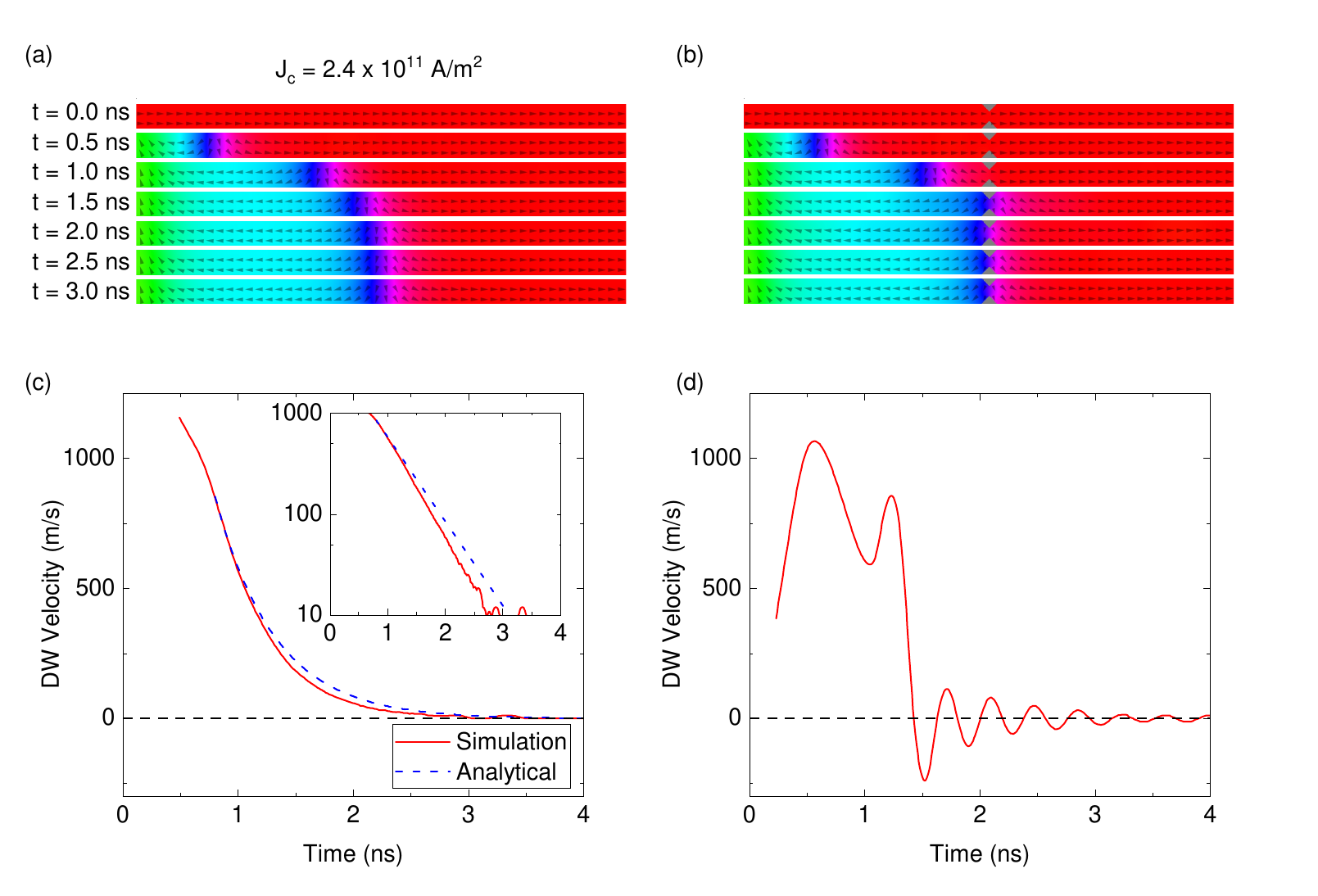}
    \caption{Micromagnetic snapshots of an isolated DW, taken every 0.5 ns from the start of the simulation in the (a) perfect and (b) notched nanostrips. The associated DW velocity as a function of simulation time is shown for the (c) perfect and (d) notched nanostrips. The inset in (c) shows the DW velocity on a logarithmic scale.}
    \label{fig:isolatedDW}
\end{figure*}

\textit{Perfect Nanostrip -} We begin with the dynamics of a single DW injected by the spin polarized charge current density mechanism mentioned above. The micromagnetic snapshots in Fig.~\ref{fig:isolatedDW}(a) (also see Supplemental Video 1 
\cite{[{See Supplemental Material at (URL to be provided by publisher) for videos depicting magnetization dynamics in the various regimes dicussed}] Supplemental}) 
show the isolated DW propagating along the nanostrip and coming to rest in the middle of the nanostrip. 
This is the point at which the total energy of the system with an isolated DW reaches a local minimum; the spin-transfer torque in the injection region is too weak to overcome the magnetostatically favored configuration where the strip is divided into two oppositely magnetized domains of equal size. 
The velocity of the isolated DW in the micromagnetic simulations, shown in Fig.~\ref{fig:isolatedDW}(c), decays in an exponential, diffusive manner. The simulation data shows an exponential decay time scale of $\tau = 0.45$ ns. 

This diffusive motion (exponentially decaying velocity) of the isolated DW agrees with our one-dimensional analytical model (details given in Appendix \ref{app:model}) in which the DW velocity is given by 
\begin{equation}
\label{eq:dwvelocity2}
v(t)=\lambda\gamma_K\Phi_0e^{-\alpha\gamma_Kt}.
\end{equation}
Here $\lambda \approx 90$ nm is the DW width, $\gamma_K = \frac{K_\perp}{s(1+\alpha^2)}$ is a rate governed by the strength of the easy-plane anisotropy, $K_\perp$, and the spin density, $s$; $\Phi_0$ is the initial out-of-plane canting angle of the DW. Based on our material parameters our model predicts the velocity decays on a time scale $\tau = (\alpha\gamma_K)^{-1} = 0.52$ ns. The DW velocity predicted by our model, shown by the dashed blue curve in Fig. \ref{fig:isolatedDW}(c), is in good qualitative agreement with the simulation results.

\textit{Notched Nanostrip -} In the notched nanostrip, the isolated DW also experiences exponentially decaying motion. However, the motion is further complicated by an additional attractive force acting on the DW from the notch defect. The isolated DW propagates towards the notches and upon reaching the notch defect, the DW undergoes damped harmonic oscillations, as seen in Fig.~\ref{fig:isolatedDW}(d), eventually becoming pinned at the defect in the center of the nanostrip (see Fig.~\ref{fig:isolatedDW}(b) and Supplemental Video 2). These oscillations of the DW about the center of a notch potential have previously been observed experimentally \cite{Saitoh2004}.

We conclude that both the perfect and notched nanostrips exhibit qualitatively similar behavior in the sense that the isolated DW is unable to propagate beyond the center of the nanostrip, either as a result of diffusive motion or DW pinning.


\subsection{Weakly Interacting Domain Wall Train}\label{sec:weaktrain}

\begin{figure*}[ht]
    \centering
    \includegraphics[width=\textwidth,trim = 0in 0in 0in 1in]{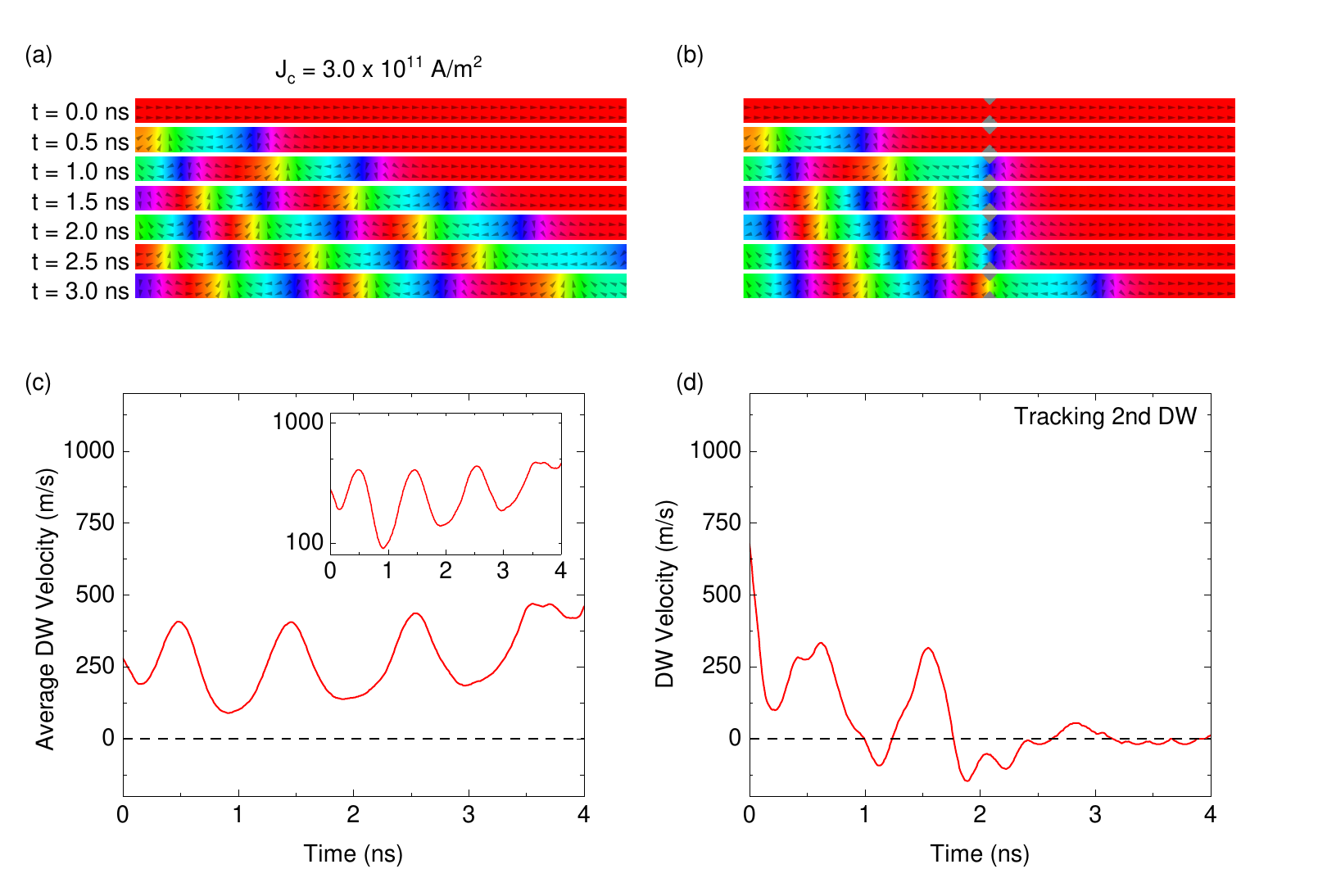}
    \caption{Micromagnetic snapshots of a weakly interacting DW train in the (a) perfect nanostrip and (b) notched nanostrip. In the notched nanostrip, note the momentary pinning of the first DW and the subsequent pinning of the DW train. The average DW velocity as a function of simulation time for the (c) perfect nanostrip and (d) the second DW in the train in the notched nanostrip. The inset in (c) shows the average DW velocity on a logarithmic scale.}
    \label{fig:weakDWtrain}
\end{figure*}

Next we consider the motion of a weakly interacting DW train. By increasing the drive charge current density to $J_c = 3.0 \times 10^{11}$ A/m$^2$, multiple DWs can now be injected into the nanostrips, shown in Figs.~\ref{fig:weakDWtrain}(a,b) and Supplemental Videos 3 and 4. 

\textit{Perfect Nanostrip - } In the perfect nanostrip the DWs individually continue to undergo exponentially decaying motion that is consistent with the behavior predicted by our model. This is shown by the DW velocity averaged across multiple DWs in the simulation in Fig.~\ref{fig:weakDWtrain}(c) (inset shows average DW velocity on a logarithmic scale). 

As multiple DWs are injected into the nanostrip, they interact in a repulsive manner as a result of the homochirality of the DWs \cite{Kruger2012,Jang2012}. These inter-DW interactions, similar to Coulomb repulsion, become responsible for the movement of the DW train past the middle of the nanostrip. Beyond the center point of the nanostrip, the repulsive interactions are aided by the DWs being attracted to the end of the nanostrip, where they are then annihilated at the sink.

\textit{Notched Nanostrip -} In the notched nanostrip we also observe repulsive DW interactions, but the dynamics is now further complicated due to the notch defect. For $J_c = 3.0 \times 10^{11}$ A/m$^2$, the first injected DW propagates towards and is pinned at the notch defect, similar to that of an isolated DW. Meanwhile, additional DWs continue to be injected into the nanostrip, allowing for a series of DWs to build up behind the notch defect. This build-up eventually pushes the first DW through the pinning site, as seen in the micromagnetic snapshots in Fig.~\ref{fig:weakDWtrain}(b). 

Once the leading DW has been pushed through the notch defect, it is attracted to the end of the nanostrip and annihilated. The second DW in the train is pushed along via the inter-DW interactions and then pinned at the notch defect. The corresponding DW velocity for this specific DW is shown in Fig.~\ref{fig:weakDWtrain}(d). At this point, no additional DWs can be injected into the strip for the remainder of the simulation. The system reaches a steady state where the energy barrier to nucleate DWs is higher than the energy provided by current-induced spin-transfer torque. 

We emphasize that the results in Figs.~\ref{fig:weakDWtrain}(b)(d) and Supplemental Video 4 do not show ``fluid-like'' dynamics -- i.e., the spin transport is not hydrodynamic. Rather than flowing past the constriction as a fluid would, the spin transport is halted at the defect; the spin-transfer torque in the injection region is too weak to nucleate additional DWs and propel the train past the defect. Thus, at low drives, DW pinning provides a natural way to understand the interaction of easy-plane precessional spin transport with defects.



\subsection{Moderately Interacting Domain Wall Train}\label{sec:moderatetrain}

We now increase the charge current density to $J_c = 4.0 \times 10^{11}$ A/m$^2$ and observe the effect of increased DW density on pinning.

\begin{figure*}[ht]
    \centering
    \includegraphics[width=\textwidth,trim = 0in 0in 0in 1in]{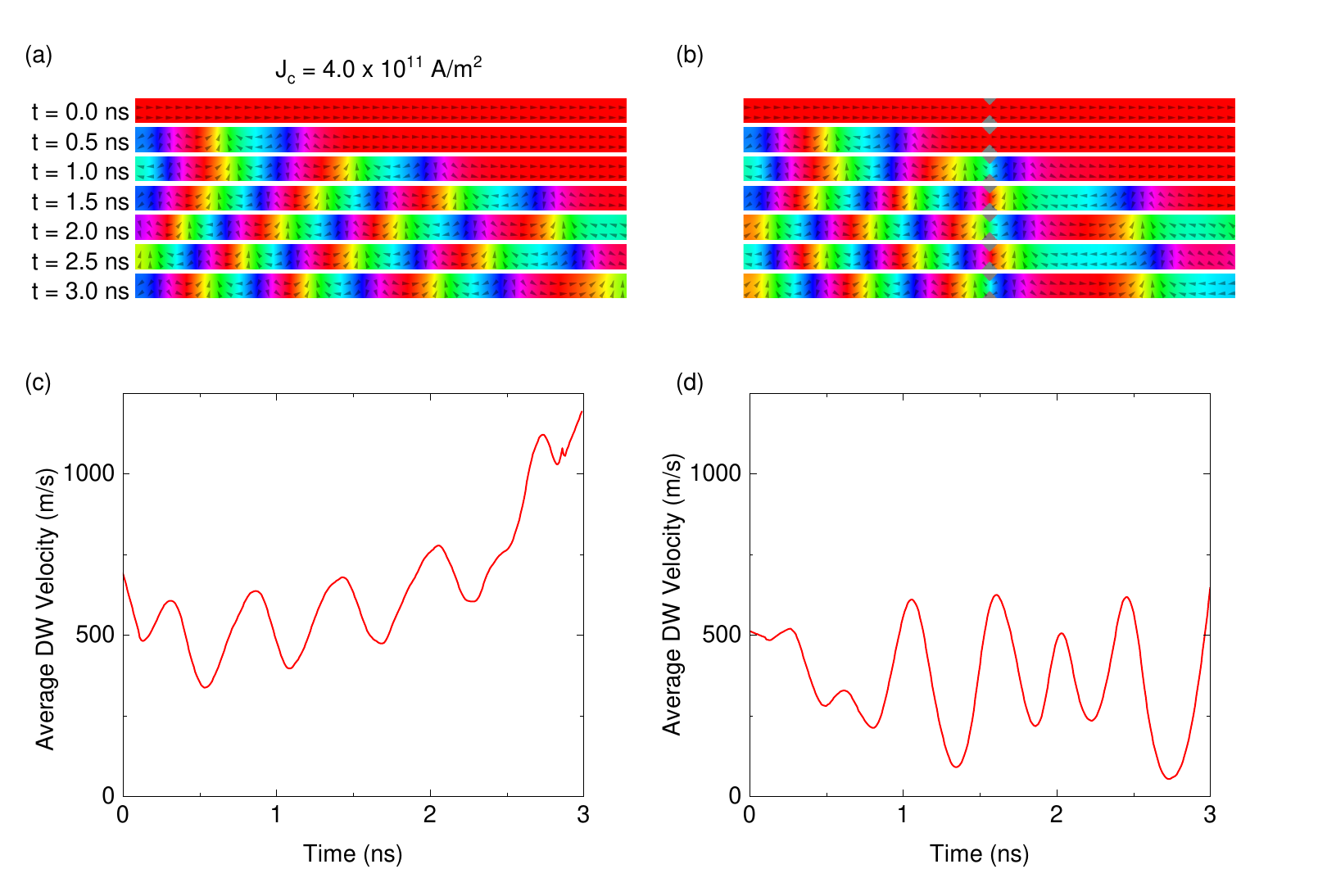}
    \caption{Micromagnetic snapshots of a weakly interacting DW train in the (a) perfect nanostrip and (b) notched nanostrip with the DW interactions are strong enough to overcome the pinning potential. The associated average DW velocity is shown for the (c) perfect and (d) notched nanostrips.}
    \label{fig:moderateDWtrain}
\end{figure*}

\textit{Perfect Nanostrip -} The increased current density yields behavior similar to that discussed in Sec.~\ref{sec:weaktrain} for the perfect nanostrip. The density of the DW train increases as more DWs can be injected into the nanostrip, see Fig.~\ref{fig:moderateDWtrain}(a) and Supplemental Video 5. The average DW velocity, shown in Fig.~\ref{fig:moderateDWtrain}(c), shows a periodic behavior as the DWs are pushed away from trailing walls and slow down as they approach the next DW in the train. As a result of the increased density of DWs, and thus stronger repulsion between neighboring DWs, the average velocity is higher than in the case where $J_c = 3.0 \times 10^{11}$ A/m$^2$ (see Sec.~\ref{sec:weaktrain} and Figs.~\ref{fig:weakDWtrain}(a,c)). The continuous motion of the DW train shown in Fig.~\ref{fig:moderateDWtrain}(a,c) is beginning to approach the fluid-like regime. 

\textit{Notched Nanostrip - } At $J_c = 4.0 \times 10^{11}$ A/m$^2$, the pinning of the DW train disappears as a result of the stronger inter-DW interactions. The DWs are still impeded by the notch defect (Fig.~\ref{fig:moderateDWtrain}(b), Supplemental Video 6), evident by the reduction in average DW velocity in Fig.~\ref{fig:moderateDWtrain}(d) when compared with the perfect nanostrip in Fig.~\ref{fig:moderateDWtrain}(c). However, they are pushed through before they can be pinned entirely, allowing for the DW train to move continuously throughout the nanostrip. 

We observe that as the driving current density is increased, the density of the DWs increases. The increased DW density allows for individual DWs in the train to be less susceptible to pinning as a result of the stronger mutual repulsion between the homochiral DWs. The overall behavior of the magnetization in the nanostrips starts to approach that of fluid-like dynamics. This point is further verified by increasing the current density to higher values, as discussed in the next section. 


\subsection{Strongly Interacting Domain Wall Train}\label{sec:condensedtrain}

\begin{figure*}[ht]
    \centering
    \includegraphics[width=\textwidth,trim = 0in 0in 0in 1in]{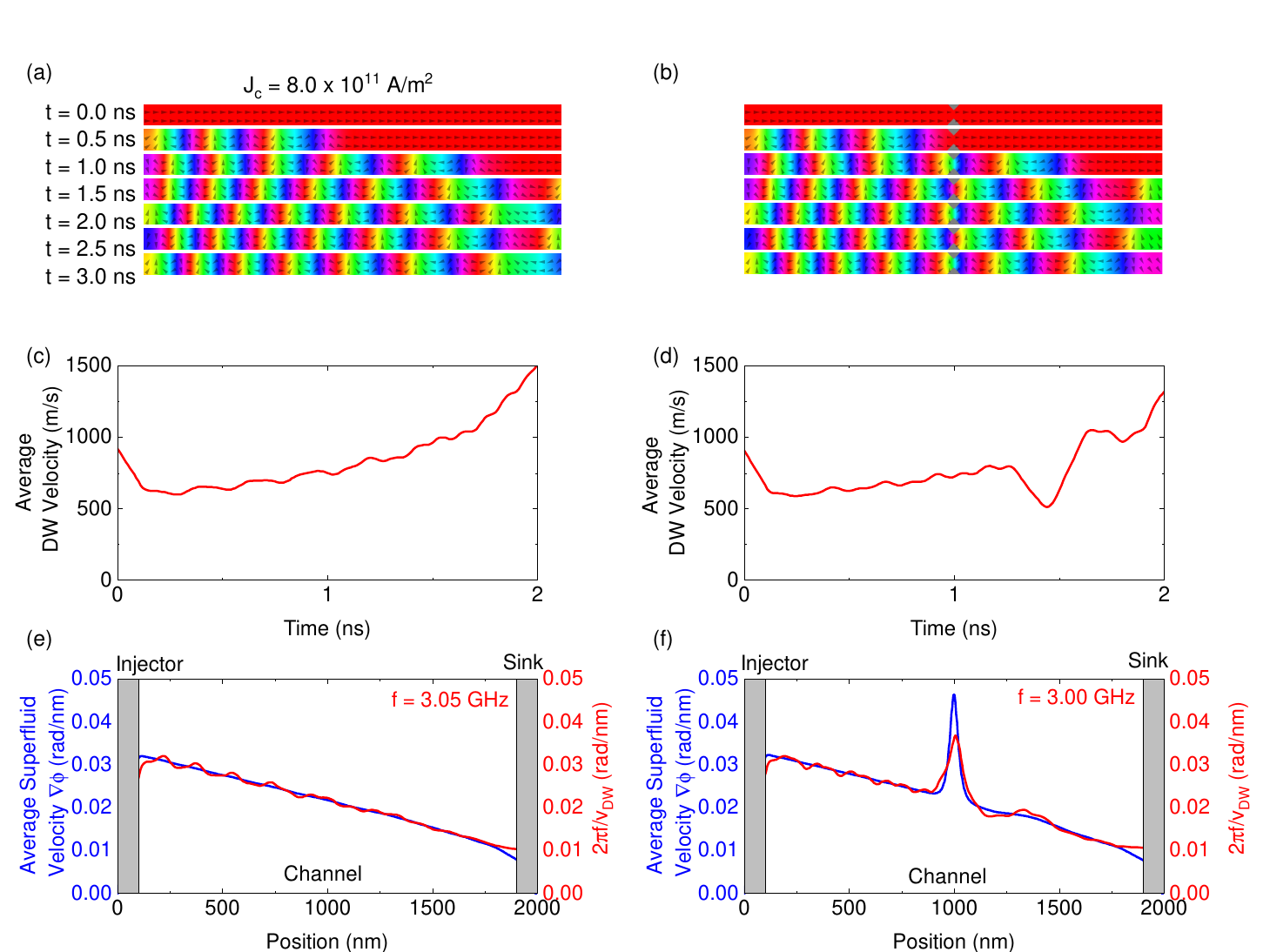}
    \caption{(a), (b) Micromagnetic snapshots of the densely packed DW train that resembles superfluid-like spin transport. (c), (d) The average DW velocity as a function of simulation time for the (c) perfect and (d) notched nanostrips. (e), (f) Time-averaged superfluid velocity and equivalent DW velocity, computed via Eq.~\ref{eq:SF-DW}, as a function of DW position for the (e) perfect and (f) notched nanostrips.}
    \label{fig:strongDWtrain}
\end{figure*}

Finally, we examine the regime of a strongly interacting, dense DW train at $J_c = 8.0 \times 10^{11}$ A/m$^2$. Micromagnetic snapshots are shown in Figs.~\ref{fig:strongDWtrain}(a,b), as well as Supplemental Videos 7 and 8, for the two geometries.

\textit{Perfect Nanostrip -}  In the perfect nanostrip, the DW train has condensed to the point that the DW separation distance is comparable to the individual DW width $\sim 100$ nm. At this point, the overall dynamics of the nanostrip begins to resemble that of superfluid-like spin transport~\cite{Sonin2010,Chen2014,Takei2014,Skarsvag2015,Tserkovnyak2018,Iacocca2017PRBa,Schneider2021} in the sense that the magnetization at a fixed position is precessing uniformly with simulation time. The average DW velocity, shown in Fig.~\ref{fig:strongDWtrain}(c), no longer shows signs of the exponential decay of an individual DW. In fact, the DW velocity continues to increase as the DW traverses the strip. As they propagate further, the DW train begins to separate and individual DWs are attracted to the end of the strip where they are eventually annihilated. 

\textit{Notched Nanostrip -} In the notched nanostrip, the inter-DW interactions of the dense train have become strong enough to overcome the pinning potential well. As the DWs impinge on the notch defect, the pinning potential reduces the speed of the DW train momentarily, before the DWs are pushed through and become attracted to the end of the strip and speed up again. The reduction in DW velocity from the notch defect can be seen clearly in Fig.~\ref{fig:strongDWtrain}(d). We also note the remarkable similarity in average DW velocity between the perfect and notched nanostrips up to the point of the notch defect. 

\textit{Convergence to Fluid-like Regime -} Our simulation results on the motion of a train of DWs showed pinning behavior present at lower $J_c$ in notched nanostrips. At sufficiently high $J_c$, the pinning behavior vanishes and the DW perspective begins to converge with the hydrodynamic one. To show further agreement with the established hydrodynamic model, we relate the DW velocity to the conventional superfluid velocity $\nabla\phi$ (where in the hydrodynamic model the spin current $J_s \propto \nabla\phi$ \cite{Sonin2010}) through the following relationship:
\begin{equation}
    \nabla\phi = \frac{2\pi f}{v_{DW}}.
    \label{eq:SF-DW}
\end{equation}
Here $\phi$ is the in-plane angle the magnetization makes with the $\hat{x}$ axis, $\nabla\phi$ is the spatial gradient of $\phi$ (given in rad/nm), $f$ is the precessional frequency of the magnetization, and $v_{DW}$ is the average DW velocity. 

We compute time-averaged $\nabla\phi$ directly (blue line) at each cell after reaching a steady state and compare it with the equivalent quantity using the average DW velocity (red line) in Fig.~\ref{fig:strongDWtrain}(e) and Fig.~\ref{fig:strongDWtrain}(f). We first note the mostly linear decay of $\nabla\phi$ in the channel, indicating that we are indeed simulating easy-plane spin transport in the fluid-like regime at $J_c = 8.0 \times 10^{11}$ A/m$^2$. In this fluid-like regime, we find an excellent quantitative agreement between the hydrodynamic and DW perspectives for both the perfect and the notched nanostrips. This agreements confirms that a densely packed DW train behaves as a ``fluid'' and convergences with the hydrodynamic model. 

In the notched nanostrips, the rapid increase in $\nabla\phi$ resulting from the constriction created by the notches is recreated well by our DW perspective. This increase in $\nabla\phi$, akin to throttling of a fluid, is also in great quantitative agreement with the DW perspective: The increase in $\nabla\phi$ corresponding with a reduction in DW velocity as the DWs propagate through the notch defect. 


\subsection{Consequences for Practical Applications}\label{sec:applications}

\begin{figure}[h]
    \centering
    \includegraphics[width=\columnwidth, trim = 1in 0.5in 0in 0.5in]{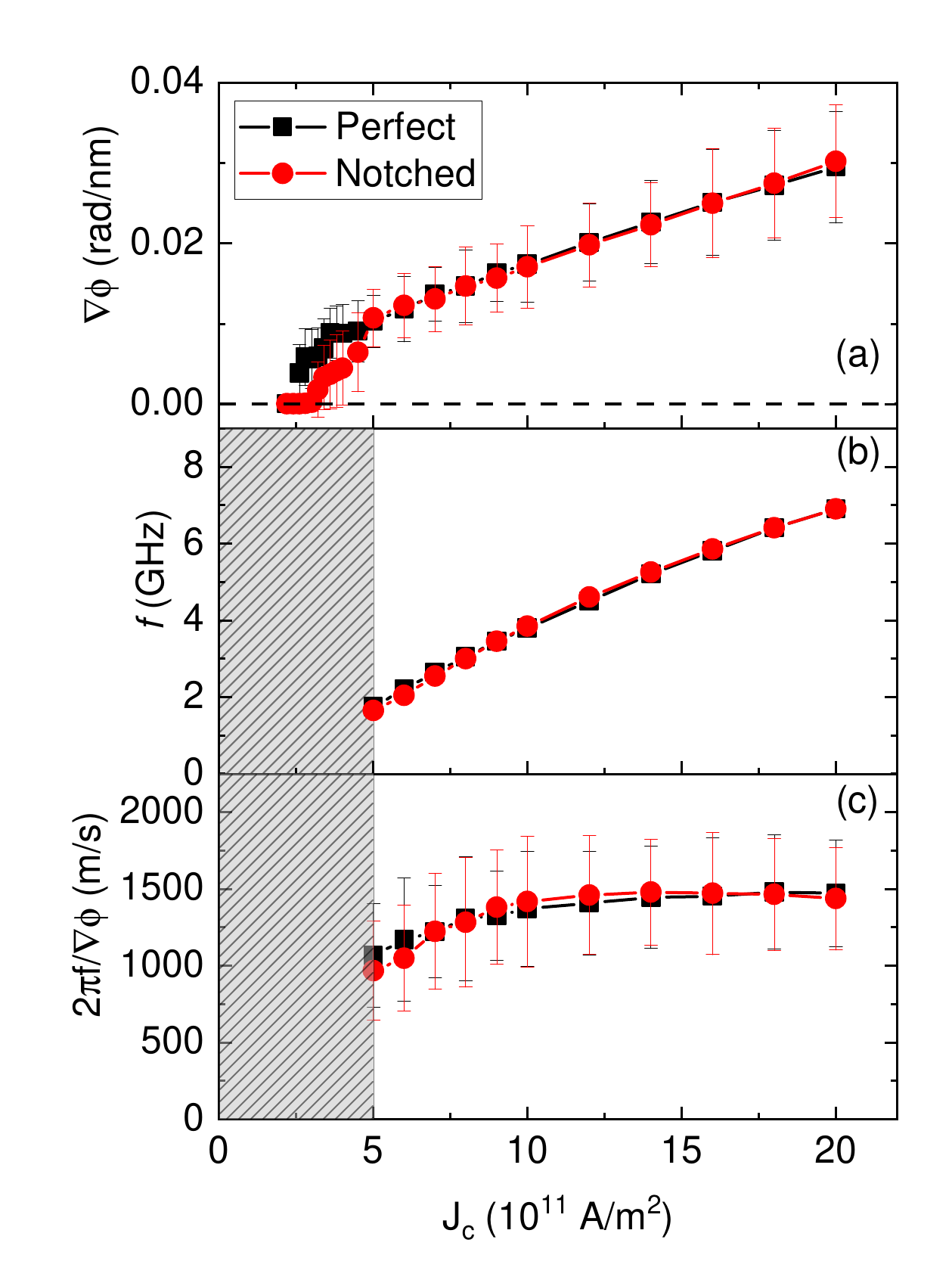}
    \caption{(a) Time-averaged superfluid velocity at $x = 1500$ nm as a function of driving current density $J_c$ for the perfect (black squares) and notched (red circles) nanostrips. The error bars indicate the standard deviation. (b) Precessional frequency of the magnetization. (c) Equivalent DW velocity computed using Eq.~\ref{eq:SF-DW} at $x = 1500$ nm}
    \label{fig:SuperfluidVelocity}
\end{figure}

We now comment on the impacts our simulation results would have on experimental realizations of easy-plane precessional dynamics. In Fig.~\ref{fig:SuperfluidVelocity}(a) we compare the time-averaged superfluid velocity $\nabla\phi$ as a function of charge current density $J_c$. The superfluid velocity shown in Fig.~\ref{fig:SuperfluidVelocity}(a) was computed at $x = 1500$ nm, beyond the location of the notch defect, for both the perfect and notched nanostrips. 

At low values of $J_c$ ($< 5\times10^{11}$ A/m$^2$), we note a difference in the superfluid velocity between the two geometries. This is a result of pinning by the notch defect, impeding individual DWs within the train. The pinning behavior disappears with increasing $J_c$ and the superfluid velocities in the two geometries become indistinguishable.  Thus, at sufficiently high $J_c$, the notch defect evidently has no effect on the global dynamics of easy-plane precession. Remarkably the pinning vanishes despite the rather large size of the defect; at their deepest point, the pair of notches occupy $60\%$ of the nanostrip's width, much larger than the typical edge roughness that results from lithographic patterning \cite{Dutta2015}. The robust transport, unaffected by such deep notches, is promising for achieving easy-plane precessional dynamics in lithographically patterned nanostrips. 

To determine the equivalent DW velocity using Eq.~\ref{eq:SF-DW}, the precessional frequency $f$ of the magnetization is determined using a fast Fourier transform on $m_x$ as a function of time along the length of the nanostrip. We limit our determination of $f$ to the fluid-like regime in which $f$ is uniform throughout the nanostrip. Precessional frequency and equivalent DW velocity as a function of $J_c$ are plotted in Fig.~\ref{fig:SuperfluidVelocity}(b) and Fig.~\ref{fig:SuperfluidVelocity}(c), respectively. The superfluid velocity $\nabla\phi$ and precessional frequency $f$ continuously increase with $J_c$ but the DW velocity saturates at $\approx 1500$ m/s. This saturation value is much higher than the typical experimentally measured value in in-plane magnetized strips \cite{Beach2005,Parkin2008,Lepadatu2017}, yet well below the maximum magnon group velocity in our system of $\approx 8000$ m/s (derived from a micromagnetically computed magnon dispersion curve), which has been suggested to be the upper limit on DW velocity \cite{Caretta2020}. Instead of being limited by the magnon group velocity, the upper bound of the DW speed in our case appears to be closer to the minimum magnon phase velocity ($\approx$2000 m/s), which previously has been shown to restrict the speed of a single transverse N\'eel DW \cite{Yan2011}. 

Our material parameters were chosen based on experimentally measured thin films of Fe$_{80}$V$_{20}$ with $\alpha = 0.006$ (see Appendix~\ref{app:PolyFeV}). This choice is in contrast to the typically chosen insulating ferrimagnetic oxide of ytrrium iron garnet (YIG) with $\alpha \sim 10^{-5} - 10^{-4}$. However, YIG is notoriously challenging to grow and integrate into practical devices, as it requires fine control of deposition parameters and high processing temperatures. FeV alloys were chosen for their low-loss magnetic properties \cite{Smith2020} and compatibility with CMOS-friendly Si substrates when deposited at room temperature~\cite{Arora2021}. Even though FeV alloys possess a damping parameter an order of magnitude larger than YIG, we were able to simulate fluid-like easy-plane spin transport at moderately achievable current densities (defined as when $\nabla\phi$ is the same for both the perfect and notched nanostrips, via Fig. \ref{fig:SuperfluidVelocity}(a)) at $J_c = 5.0 \times 10^{11}$ A/m$^2$. At lower current densities, $J_c \approx 3 \times 10^{11}$ A/m$^2$, the DW train could overcome pinning and was able to propagate throughout the entirety of the nanostrip. This would still allow for spin transport along the nanostrip (as a result of the rotating magnetization in the spin sink region) and the possibility of efficient micron-scale transmission of spin-based information. 

Our chosen method of excitation simulates a current-perpendicular-to-plane spin valve nanopillar with an out-of-plane polarizer. This is a well established technique in orthogonal spin-torque oscillators \cite{Houssameddine2007}. Thus, the simulated dynamics here in principle can be achieved using experimentally proven physics and device structures. Additionally, recent studies have pointed to the possibility of in-plane magnetized films producing an out-of-plane spin torque \cite{Baek2018,MacNeill2017}. This out-of-plane spin-orbit torque could prove to be a viable method of exciting easy-plane precessional dynamics as it would eliminate the need for complicated fabrication of nanopillar spin valves. However, it is unclear at this time if this torque would be strong enough to drive the easy-plane precession dynamics simulated here. 

It is worth pointing out that while our simulations were performed at zero temperature, experimental attempts at achieving easy-plane precessional dynamics will be done at finite temperatures. Finite temperatures allow for the emergence of diffusive thermal magnon transport, which could couple to the easy-plane spin transport and provide another avenue for dissipation that is not captured by the Gilbert damping parameter \cite{Sonin2019}. In our zero-temperature simulations, there are no thermal magnons that could give rise to the additional non-Gilbert dissipation. While possible dissipation pathways via thermal magnons are beyond the scope of this present work, future studies employing finite-temperature micromagnetic simulations may give insights into such dissipation in easy-plane spin transport.

\section{Conclusion}

We performed micromagnetic simulations on the interaction of homochiral DW transport via easy-plane precession in synthetic antiferromagnet nanostrips with and without a notch defect. We observed the diffusive motion of an isolated DW and subsequent pinning at the notch defect at low $J_c$. With increasing $J_c$ multiple DWs are injected into the nanostrip, and we observed the crossover to a fluid-like, densely packed DW train. The densely packed DW train in notched nanostrips is robust to edge defects and shows no difference to the perfect nanostrips in the fluid-like regime. Our simulations, with material parameters taken directly from experimentally measured metallic ferromagnets, demonstrate promise for an experimental realization of easy-plane precession at reasonable current densities for efficient micron-scale spin transport. 


\section{Acknowledgements}
D.A.S., L.C., F.R.-D., and S.E. acknowledge support by NSF Grant No. DMR-2003914. S.T. acknowledges support by CUNY Research Foundation Project $\#$ 90922-07 10 and PSC-CUNY Research Award Program $\#$ 63515-00 51. M.S. acknowledges support by the Luther and Alice Hamlett Undergraduate Research Support Program. 


\bibliography{bibliography}


\appendix


\section{Easy-plane Precession Dynamics in Single Layer Systems}\label{app:singlelayer}

\begin{figure*}
    \centering
    \includegraphics[width=\textwidth,trim=0in 4.5in 0in 1.5in]{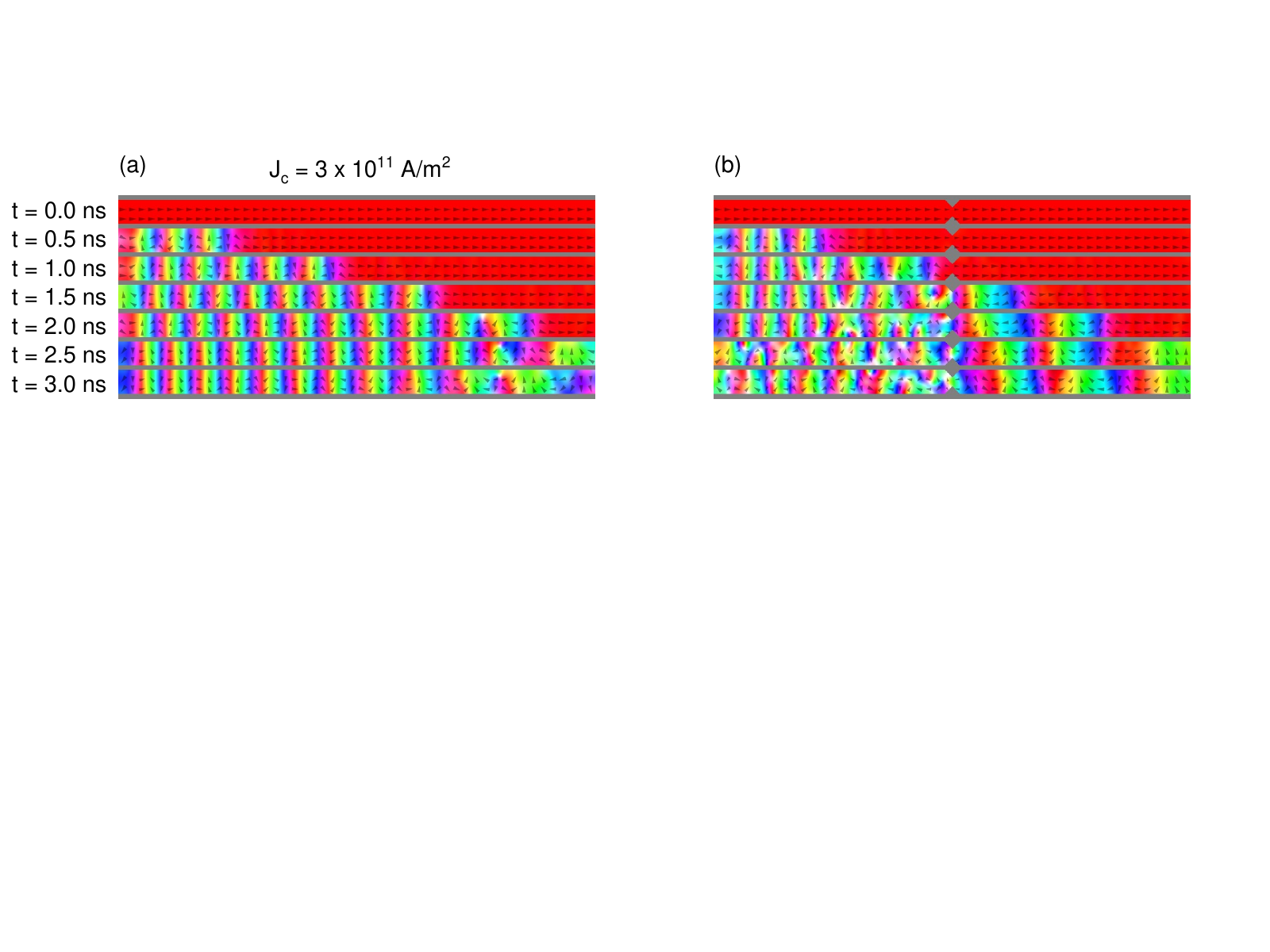}
    \caption{Micromagnetic snapshots of vortex formation in single layer (a) perfect and (b) notched nanostrips}
    \label{fig:SL-vortex}
\end{figure*}

We focused on simulating easy-plane spin transport in synthetic antiferromagnets as opposed to single layer nanostrips. In synthetic antiferromagnets, the long-range dipolar fields from one ferromagnetic layer are compensated by an adjacent second layer. This has the effect of stabilizing transverse N\'eel DWs and suppressing Walker breakdown \cite{Lepadatu2017}. Micromagnetic snapshots of phase slips via vortex formation (similar to Walker breakdown) in single layer systems are shown in Fig.~\ref{fig:SL-vortex}(a) and Fig.~\ref{fig:SL-vortex}(b) for the perfect and notched nanostrips, respectively. Supplemental Videos 9 and 10 complement the micromagnetic snapshots shown in Figs.~\ref{fig:SL-vortex}(a,b).

In the perfect nanostrip, a vortex cores begins to form at the end of the nanostrip within a DW. The vortex core then propogates against the flow of DWs. In the notched nanostrips, multiple vortex cores begin to form at the edges of the nanostrip, similar to the perfect nanostrip. The vortex fully forms off the tip of the notch defect (see Supplemental Video 10). These vortices stay in the nanostrip until they encounter a vortex with opposite core polarity upon which the pair is annihilated. 

The difference between the single layer (Fig.~\ref{fig:SL-vortex}) and synthetic antiferromagnet systems (Fig.~\ref{fig:weakDWtrain}) is striking. The formation of vortices is absent in synthetic antiferromagnet systems up to high drive current densities $J_c \gtrsim 2 \times 10^{12}$ A/m$^2$, even in notched nanostrips.


\section{Experimental Determination of Material Parameters}\label{app:PolyFeV}

The material parameter chosen for our micromagnetic simulations were similar to those of experimentally measured polycrystalline Fe$_{80}$V$_{20}$ thin films. We deposited these films using magnetron sputtering with base pressure $< 5 \times 10^{-8}$ Torr. The films were deposited on Si/SiO$_2$ substrates at room temperature with an Ar pressure of $3$ mTorr. A Ti/Cu seed layer was initially deposited to promote good adhesion to the substrate and a Ti capping layer was deposited to protect against film oxidation. Fe and V were co-sputtered from two separate targets. All material deposition rates were calibrated using x-ray reflectivity. The sample stack structure is subs./Ti(3)/Cu(3)/Fe$_{80}$V$_{20}$(2)/Ti(3) where the values in the parentheses are layer thicknesses in nm. 

To determine the magnetic properties of our films, we utilized broadband ferromagnetic resonance (FMR). The thin film sample was placed face-down on a coplanar waveguide with a maximum frequency of $36$ GHz and magnetized by an external field $H$ generated by a conventional electromagnet. The FMR spectra was acquired by fixing the microwave frequency and sweeping the magnetic field through the resonance condition. The resulting spectra is then fit with a Lorentzian derivative, from which the resonance field $H_{res}$ and half-width-at-half-maximum (HWHM) linewidth $\Delta H$ are determined for each frequency.

\begin{figure}
    \centering
    \includegraphics[width=\columnwidth]{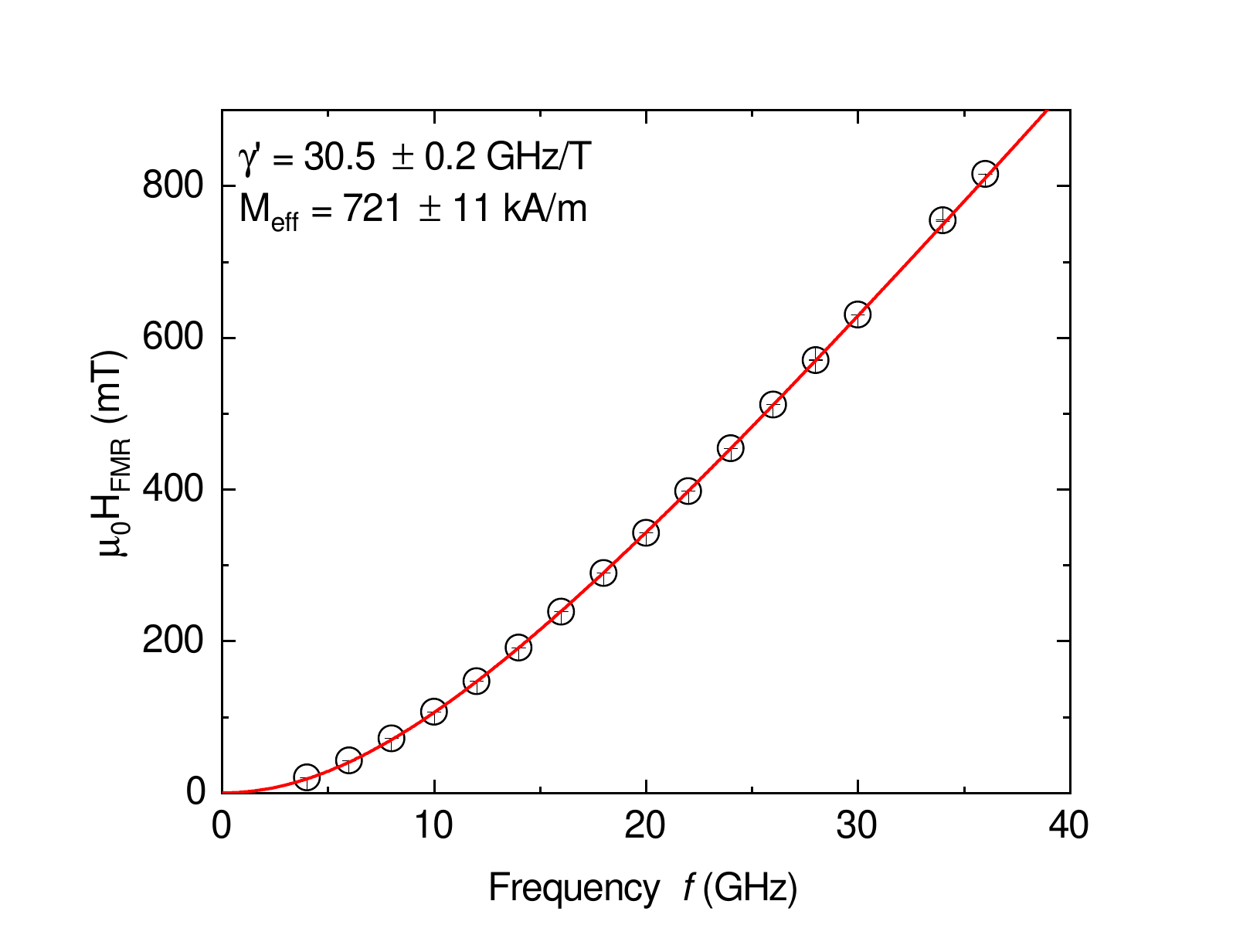}
    \caption{FMR resonance field as a function of microwave frequency. The solid line is a fit according to Eq.~\ref{eq:kittel}.}
    \label{fig:kittel}
\end{figure}

The resonance field as a function of microwave frequency is plotted in Fig.~ \ref{fig:kittel} and fit using the standard Kittel equation \cite{Kittel1948}
\begin{equation}
    f = \mu_0\gamma'\sqrt{H_{res}(H_{res}+M_{eff})},
    \label{eq:kittel}
\end{equation}

\noindent where $\gamma' = \gamma/2\pi$ is the reduced gyromagnetic ratio and $M_{eff}$ is the effective magnetization (here equal to the saturation magnetization $M_{sat}$). From this fit we determine that $\gamma' \approx 30.5$ GHz/T and $M_{eff} = 720$ kA/m.  

\begin{figure}
    \centering
    \includegraphics[width=\columnwidth]{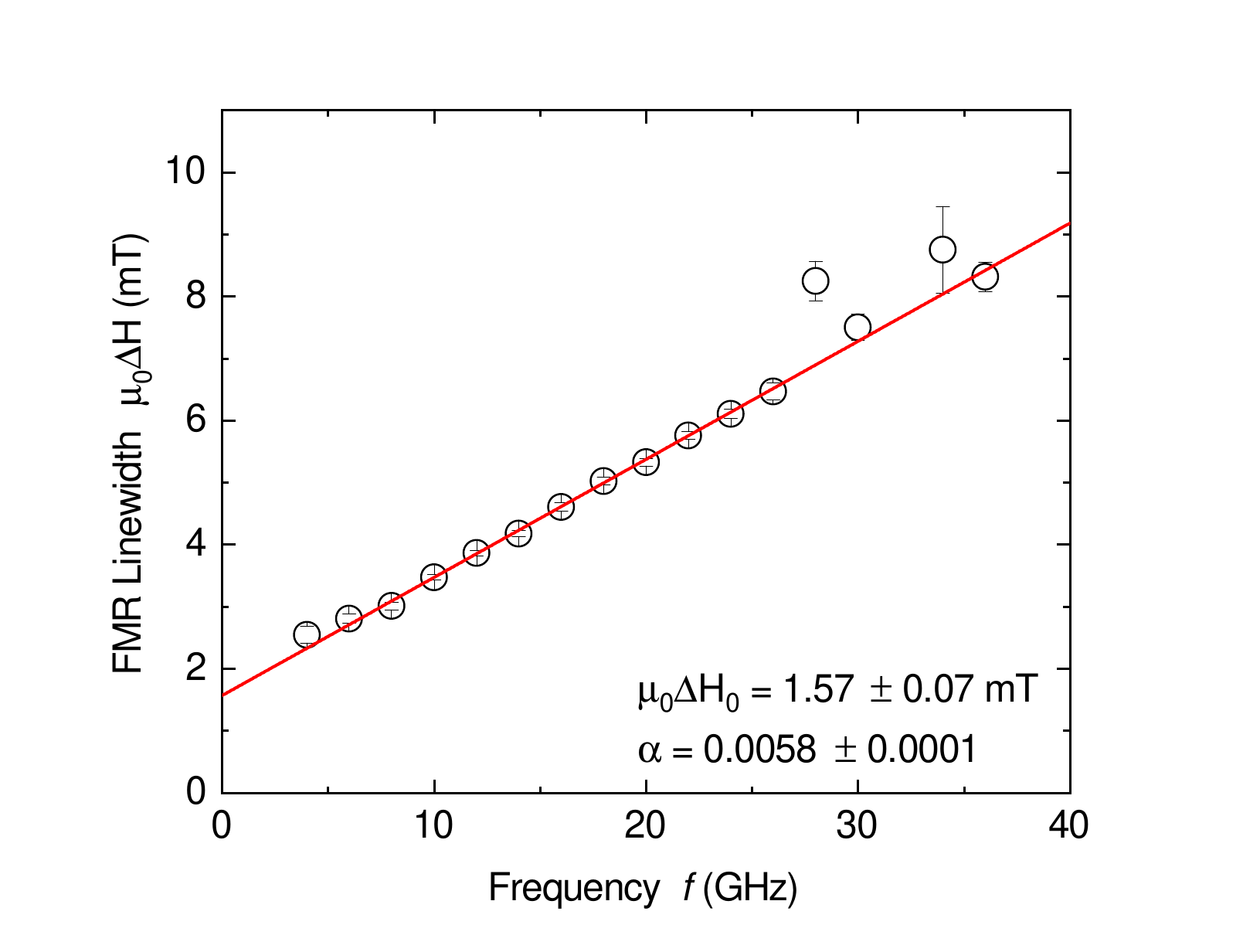}
    \caption{FMR linewidth as a function of microwave frequency. The solid line is a fit according to Eq.~\ref{eq:linewidth}}
    \label{fig:linewidth}
\end{figure}

The HWHM linewidth, plotted in Fig.~\ref{fig:linewidth}, gives insight into the magnetic relaxation of a film. By using the linear equation \cite{Heinrich2005}
\begin{equation}
    \Delta H = \Delta H_0 + \frac{\alpha}{\mu_0\gamma'}f
    \label{eq:linewidth}
\end{equation}

\noindent one can determine the Gilbert damping parameter $\alpha$ and zero frequency linewidth $\Delta H_0$. From the linear fit we deduce $\alpha = 0.006$ in our 2 nm FeV film.

\section{Analytical Model Details}\label{app:model}

The synthetic antiferromagnet (SAF) consists of two identical ferromagnetic nanostrips coupled antiferromagnetically; the nanostrips are labeled by $i=1,2$ and are modeled as quasi-one dimensional spin chains for simplicity. We adopt a coordinate system in which the SAF extends along the $x$ axis with the strip plane oriented normal to the $z$ axis. The SAF Hamiltonian can then be written as

\begin{multline}
H_0[{\boldsymbol n_i}]=\frac{1}{2}\sum_{i=1,2}\int dx\,\big[A(\partial_x{\boldsymbol n_i}(x))^2\\
+K_\perp n_{i,z}^2(x)-K_\parallel n_{i,x}^2(x)\big],
\label{eq:h0}
\end{multline}
where $A$ is the exchange stiffness, $K_\perp>0$ is the easy-plane anisotropy (with the hard axis along the $z$ axis), $K_\parallel>0$ is the easy-axis anisotropy along the $x$ axis, and the unit vector field ${\boldsymbol n_i}(x)$ points parallel to the saturated local spin density ${\boldsymbol s_i}(x)=s{\boldsymbol n_i}(x)$. 
Finally, we assume the two ferromagnets couple through an isotropic antiferromagnetic exchange interaction described by the Hamiltonian,
\begin{equation}
\label{eq:hc}
H_c[{\boldsymbol n_i}]=\eta\int dx\,{\boldsymbol n}_{1}(x)\cdot{\boldsymbol n}_{2}(x).
\end{equation}

For low enough excitation energies, DW dynamics in each layer can be described sufficiently in terms of two ``soft'' variables: the DW position $X_i(t)$ and the spin canting angle out of the easy ($xy$) plane $\phi_i(x,t)=\phi_i(t)$, the latter of which is taken to be uniform along the strip. Focusing exclusively on DWs of the N\'eel type, an appropriate parametrization for ${\boldsymbol n}_i$ in terms of these soft modes is given by~\cite{Schryer1974}, 
\begin{equation}
\label{eq:n}
{\boldsymbol n}_i(x,t)=\left(\begin{array}{c}b_i\tanh\left(\frac{x-X_i(t)}{\lambda}\right)\\b_i\chi_i\sech\left(\frac{x-X_i(t)}{\lambda}\right)\cos\phi_i(t)\\\sech\left(\frac{x-X_i(t)}{\lambda}\right)\sin\phi_i(t)\end{array}\right),
\end{equation}
where $\lambda=\sqrt{A/K_\parallel}$ is the DW width, $b_i=+1$ ($b_i=-1$) corresponds to tail-to-tail (head-to-head) DW, and $\chi_i=\pm1$ is the chirality of the DW. We hereafter fix $\chi_i=1$.

Reduced DW dynamics in terms of the soft variables can be obtained by first inserting Eq.~\eqref{eq:n} into the Landau-Lifshitz-Gilbert equation, 
\begin{equation}
\label{eq:llg}
\dot{\boldsymbol n}_i=\frac{1}{s}{\boldsymbol n}_i\times\left(-\frac{\delta H}{\delta{\boldsymbol n}_i}\right)-\alpha{\boldsymbol n}_i\times\dot{\boldsymbol n}_i\ ,
\end{equation}
\textemdash~$\alpha$ is the Gilbert parameter~\textemdash~and integrating out the irrelevant fast-oscillating modes by performing a spatial average over Eq.~\eqref{eq:llg}~\cite{Tretiakov2008}. The resulting equations are a coupled dynamics for the DWs in the two ferromagnetic nanostrips,
\begin{align}
\label{eq:first}
\left(\begin{array}{c}\dot X_1\\\dot\phi_1\end{array}\right)&=\frac{1}{2(1+\alpha^2)}\left(\begin{array}{cc}\alpha\lambda&-1\\1&\frac{\alpha}{\lambda}\end{array}\right)\left(\begin{array}{c}F_X\\F_{\phi}\end{array}\right),\\
\label{eq:second}
\left(\begin{array}{c}\dot X_2\\\dot\phi_2\end{array}\right)&=\frac{1}{2(1+\alpha^2)}\left(\begin{array}{cc}\alpha\lambda&-1\\1&\frac{\alpha}{\lambda}\end{array}\right)\left(\begin{array}{c}-F_X\\F_{\phi}\end{array}\right),
\end{align}
where the force terms read
\begin{align}
F_X&=\frac{2\eta}{s}\left(\frac{\xi}{\sinh^2\xi}-\coth\xi\right)\nonumber\\
&\qquad +\frac{2\eta}{s}\left(\frac{1-\xi\coth\xi}{\sinh\xi}\right)\cos(\phi_1+\phi_2),\\
F_{\phi}&=-\frac{\lambda K_\perp}{s}\sin(2\phi_1)-\frac{2\lambda\eta}{s}\frac{\xi}{\sinh\xi}\sin(\phi_1+\phi_2),
\end{align}
with $\xi\equiv(X_1-X_2)/\lambda$. For zero interlayer coupling, these equations reduce to the dynamics of two decoupled ferromagnetic DWs, as expected. 

Let us now consider the dynamics of a single SAF DW following its injection through the above-described spin-transfer torque mechanism. The injection process may result in differences in the positions and/or canting angles of the two constituent ferromagnetic DWs. Here, we focus on the limit of strong interlayer coupling and strong easy-plane anisotropy such that the injected DW obeys $X_1\approx X_2$ and $\phi_i\ll1$. 

Upon linearizing Eqs.~\eqref{eq:first} and \eqref{eq:second} with respect to $\xi\ll1$ and $\phi_i\ll1$, the center-of-mass coordinates [$\Xi\equiv(X_1+X_2)/2\lambda$ and $\Phi\equiv(\phi_1+\phi_2)/2$] and the relative coordinates ($\xi$ and $\varphi\equiv\phi_1-\phi_2$) decouple, and we arrive at
\begin{align}
\label{eq:com}
\left(\begin{array}{c}\dot\Xi\\\dot\Phi\end{array}\right)&=\left(\begin{array}{cc}0&\gamma_K\\0&-\alpha\gamma_K\end{array}\right)\left(\begin{array}{c}\Xi\\\Phi\end{array}\right)\ ,\\
\label{eq:rel}
\left(\begin{array}{c}\dot\xi\\\dot\varphi\end{array}\right)&=\left(\begin{array}{cc}-\alpha \gamma_\eta&\gamma_K\\-\gamma_\eta&-\alpha\gamma_K\end{array}\right)\left(\begin{array}{c}\xi\\\varphi\end{array}\right)\ ,
\end{align}
where
\begin{equation}
\label{eq:rates}
\gamma_\eta=\frac{2\eta}{s(1+\alpha^2)}\ ,\ \ \ \gamma_K=\frac{K_\perp}{s(1+\alpha^2)}\ .
\end{equation}
Equation~\eqref{eq:rates} are rates determined by the interlayer exchange and easy-plane anisotropy, respectively. 

The dynamics of the relative coordinates \eqref{eq:rel} shows that small mismatches in DW positions and canting angles between the top and bottom layers at the time of injection decay on a time scale $[\alpha(\gamma_\eta+\gamma_K)]^{-1}$. In the limit of very strong interlayer coupling, i.e., $\gamma_\eta\gg\gamma_K$, these interlayer mismatches decay on a very short time scale after injection and may effectively be ignored in the DW analysis. 

Now focusing on the center-of-mass dynamics~\eqref{eq:com}, the closed equation for $\Phi(t)$ may be solved straightforwardly giving
\begin{equation}
\Phi(t)=\Phi_0e^{-\alpha\gamma_Kt}\ ,
\end{equation}
Inserting this result into the equation for the DW velocity, we find that the velocity decays from its initial value over the time scale $\gamma_K^{-1}$, i.e.,
\begin{equation}
\label{eq:dwvelocity}
v(t)\equiv\lambda\dot\Xi(t)=\lambda\gamma_K\Phi_0e^{-\alpha\gamma_Kt}\ .
\end{equation}
The rate of DW velocity attenuation is governed by the easy-plane anisotropy, i.e., $\gamma_K$. Therefore, in the limit of strong interlayer coupling $\gamma_\eta\gg\gamma_K$, the velocity decays on a time scale much greater than the time scale governing the decay of the DW's internal mismatch. 




\end{document}